


\def\boringfonts{y}  
\input harvmac

\def\fonttest{y}
\ifx\boringfonts\fonttest\else

\fi

\hyphenation{anom-aly anom-alies coun-ter-term coun-ter-terms
dif-feo-mor-phism dif-fer-en-tial super-dif-fer-en-tial dif-fer-en-tials
super-dif-fer-en-tials reparam-etrize param-etrize reparam-etriza-tion}


%
%
%
\newwrite\tocfile\global\newcount\tocno\global\tocno=1
\ifx\bigans\answ \def\tocline#1{\hbox to 320pt{\hbox to 45pt{}#1}}
\else\def\tocline#1{\line{#1}}\fi
\def\toclead{\leaders\hbox to 1em{\hss.\hss}\hfill}
\def\tnewsec#1#2{\xdef #1{\the\secno}\newsec{#2}
\ifnum\tocno=1\immediate\openout\tocfile=toc.tmp\fi\global\advance\tocno
by1
{\let\the=0\edef\next{\write\tocfile{\medskip\tocline{\secsym\ #2\toclead\the
\count0}\smallskip}}\next}
}
\def\tnewsubsec#1#2{\xdef #1{\the\secno.\the\subsecno}\subsec{#2}
\ifnum\tocno=1\immediate\openout\tocfile=toc.tmp\fi\global\advance\tocno
by1
{\let\the=0\edef\next{\write\tocfile{\tocline{ \ \secsym\the\subsecno\
#2\toclead\the\count0}}}\next}
}
\def\tappendix#1#2#3{\xdef #1{#2.}\appendix{#2}{#3}
\ifnum\tocno=1\immediate\openout\tocfile=toc.tmp\fi\global\advance\tocno
by1
{\let\the=0\edef\next{\write\tocfile{\tocline{ \ #2.
#3\toclead\the\count0}}}\next}
}
%
%

%
%
%
%

\gdef\prlmode{F}
%
%
%
\let\narrowequiv=\equiv
\def\equiv{\;\narrowequiv\;}
\let\narrowtilde=\tilde
\def\tilde{\widetilde}
\fontdimen16\tensy=2.7pt\fontdimen17\tensy=2.7pt 



%
\def\ga{\gamma}
\def\la{\lambda}
\def\ep{\epsilon}

\def\dl{\delta}
%
%

\def\CP{{\cal P}}
\def\CO{{\cal O}}

\def\CE{{\cal E}}

%
%
%
\def\boxit#1#2{
        $$\vcenter{\vbox{\hrule\hbox{\vrule\kern3pt\vbox{\kern3pt
	\hbox to #1truein{\hsize=#1truein\vbox{#2}}\kern3pt}\kern3pt\vrule}
        \hrule}}$$
}


%

\def\lfr#1#2{{\textstyle{#1\over#2}}} 



\def\splitexact#1#2{\mathrel{\mathop{\null{
\lower4pt\hbox{$\rightarrow$}\atop\raise4pt\hbox{$\leftarrow$}}}\limits
^{#1}_{#2}}}

%
%
\def\pa{\partial}

\def\pd#1#2{{\partial #1\over\partial #2}} 
%
%
%
%

\def\dd{\mskip 1.3mu{\rm d}\mskip .7mu} 



%
%

\def\ie{{\it i.e.}}

%
%

\ifx\boringfonts\fonttest
\font\blackboard=cmssbx10 \font\blackboards=cmssbx10 at 7pt  
\font\blackboardss=cmssbx10 at 5pt
\else
\font\blackboard=msym10 \font\blackboards=msym7   
\font\blackboardss=msym5
\fi
\newfam\black
\textfont\black=\blackboard
\scriptfont\black=\blackboards
\scriptscriptfont\black=\blackboardss


%
\ifx\boringfonts\fonttest
\font\gothic=cmssbx10 \font\gothics=cmssbx10 at 7pt  
\font\gothicss=cmssbx10 at 5pt
\else
\font\gothic=eufm10 \font\gothics=eufm7
\font\gothicss=eufm5
\fi
\newfam\gothi
\textfont\gothi=\gothic
\scriptfont\gothi=\gothics
\scriptscriptfont\gothi=\gothicss

{\catcode`\@=11\gdef\oldcal{\fam\tw@}}
\newfam\curly
\ifx\boringfonts\fonttest\else
\font\curlyfont=eusm10 \font\curlyfonts=eusm7
\font\curlyfontss=eusm5
\textfont\curly=\curlyfont
\scriptfont\curly=\curlyfonts
\scriptscriptfont\curly=\curlyfontss
\def\cal{\fam\curly\relax}
\fi
%

\ifx\boringfonts\fonttest\def\df{\bf}\else\font\df=cmssbx10\fi

\global\newcount\pnfigno \global\pnfigno=1
\newwrite\ffile
\def\pfig#1#2{Fig.~\the\pnfigno\pnfig#1{#2}}
\def\pnfig#1#2{\xdef#1{Fig. \the\pnfigno}%
\ifnum\pnfigno=1\immediate\openout\ffile=figs.tmp\fi%
\immediate\write\ffile{\noexpand\item{\noexpand#1\ }#2}%
\global\advance\pnfigno by1}

%
%
\def\tfig#1{Fig.~\the\pnfigno\xdef#1{Fig.~\the\pnfigno}\global\advance\pnfigno
by1}

%
%
%
%
\def\figI{y}
\def\ifigure#1#2#3#4{
\midinsert
\ifx\figflag\figI
 \ifx\htflag\figI
 \vbox{
  \href{file:#3}
{Click here for enlarged figure.}}
 \fi
 \vbox to #4truein{
 \vfil\centerline{\epsfysize=#4truein\epsfbox{#3}}}
\else
\vbox to .2truein{}
\fi
\narrower\narrower\noindent{\bf #1:} #2
\endinsert
}








%
%

%


\def\df{}    

\long\def\clip#1{}
\long\def\optional#1{}

\def\rhoPN{r}\def\LambdaPN{\Lambda}
\def\cmp#1{#1 }         


\def\pndate{10/94}
\long\def\suppress#1{}
\suppress{\def\boringfonts{y}  
\def\pndate{\vbox{\vskip .2truein\hbox{{\sl Running title:} Dynamic Theory of
Pearling Instability}\hbox{{\sl PACS:} 87.10, 11.17, 68.15, 87.20}
}}
\def\figflag{n}
\baselineskip=20pt
\def\ifigure#1#2#3#4{\nfig\dumfig{#2}}
} 

\def\testp{T}

\Title{\vbox{\hbox{UPR--622T}}}{\vbox{\centerline{Dynamic Theory of
Pearling Instability}
\vskip2pt\centerline{in Cylindrical Vesicles}
}}

\centerline{Philip Nelson and Thomas Powers}\smallskip
\centerline{Physics Department, University of Pennsylvania}
\centerline{Philadelphia, PA 19104 USA}
\bigskip\centerline{Udo Seifert}\smallskip
\centerline{Institut f\"ur Festk\"orperforschung, Forschungszentrum J\"ulich}
\centerline{52425 J\"ulich, Germany}
\bigskip\bigskip

We give a simple theory for recent experiments of Bar-Ziv and Moses%
\ifx\prlmode\testp{} [Phys. Rev. Lett. {\bf73} (1994)1392]\fi%
, in which tubular vesicles are excited using laser tweezers to a
``peristaltic'' state. Considering the hydrodynamics of a bilayer
membrane under tension, {\df
we reproduce some of the qualitative behavior seen and}
find a value for the wavelength of
the instability in terms of independently measured material
parameters, in rough agreement with the experimental values.

\ifx\prlmode\testp
\fi
\ifx\answ\bigans \else\noblackbox\fi
\Date{\pndate}\noblackbox                 


\def\laz{\lambda_0}
\def\ux{u_k}
\def\micron{$\mu$m}


\def\max{_{\rm max}} \def\crit{_{\rm crit}}

\def\cmmt{cm$^{-2}$}
\def\erg{erg\thinspace}

\def\eem#1{\cdot 10^{-#1}}
\def\dvt{\narrowtilde v_+-\narrowtilde v_-}

\hfuzz=3truept

\lref\EvHe{E. Evans, Biophys. J. {\bf14} (1974) 923; W. Helfrich, \cmp{%
``Blocked lipid exchange in bilayers and its
possible influence on the shape of vesicles,''} Z. Naturforsch.
{\bf29C} (1974) 510.}
\lref\Evansc{E. Evans, Biophys. J. {\bf14} (1974) 923; Biophys. J.
{\bf30} (1980) 265.}%
\lref\SBZ{S. Svetina, M.B. Brumen, B. Zeks, ``Lipid bilayer elasticity
and the bilayer couple interpretation of red cell shape
transformations and lysis,'' Studia Biophys, {\bf110} (1985) 177.}%
\lref\Evansa{E. Evans, ``Entropy-driven tension in vesicle membranes
and unbinding of adherent vesicles,'' Langmuir {\bf7} (1991) 1900.}
\lref\Evansb{E. Evans, A, Yeung, R. Waugh, and J. Song, \cmp{``Dynamic
coupling and nonlocal curvature elasticity in bilayer membranes,''} in
{\sl Structure and conformation of amphiphilic membranes,} ed. R.
Lipowsky {\it et al} (Springer, 1992).}%
\lref\SeLa{U. Seifert and S. Langer, \cmp{``Viscous modes of bilayer
membranes,''} Europh. Lett. {\bf23}(1993)71; 
\cmp{``Hydrodynamics of membranes: the bilayer aspect and adhesion,''}
Biophys. Chem. {\bf49} (1994) 13.}
\lref\MiaSeif{U. Seifert, K. Berndl, and R. Lipowsky, \cmp{``Shape
transformation of vesicles,''} Phys. Rev. {\bf A44} (1991) 1182; 
L. Miao {\it et al.}, \cmp{``Equilibrium budding and
vesiculation in the curvature model of fluid membranes,''} Phys. Rev.
{\bf A43} (1991) 6843} 
\lref\MSWD{L. Miao {\it et al.}, \cmp{``Budding transitions of
fluid-bilayer vesicles,''} Phys. Rev. {\bf E49} (1994) 5389.}%
\lref\Miaoa{L. Miao {\it et al.}, ``Equilibrium budding and
vesiculation in the curvature model of fluid membranes,'' Phys. Rev.
{\bf A43} (1991) 6843.}
\lref\Seifert{U. Seifert, K. Berndl, and R. Lipowsky, ``Shape transformation
of vesicles,'' Phys. Rev. {\bf A44} (1991) 1182.}
\lref\Bouasse{H. Bouasse, {\sl Cours de math\'ematiques
g\'en\'erales} (Paris, Delagrave, 1911).}
\lref\DeHeb{H. Deuling and W. Helfrich, \cmp{``The curvature elasticity of
fluid membranes: a catalog of vesicle shapes,''} J. Phys. (Paris) {\bf
37} (1976) 1335.}
\lref\ShSi{M. Sheetz and S. Singer, ``Biological membranes as bilayer
couples,'' Proc. Nat. Acad. Sci. USA {\bf 71} (1974) 4457.}
\lref\DeHea{H. Deuling and W. Helfrich, \cmp{``A theoretical explanation
for the myelin shapes of red blood cells,''} Blood Cells {\bf3} (1977)
713.}%
\lref\Miaob{L. Miao {\it et al.}, ``Budding transitions of
fluid-bilayer vesicles,'' Phys. Rev. {\bf E49} (1994) 5389.}
\lref\Bozic{B. Bozic {\it et al.}, ``Role of lamellar membrane
structure in tether formation from bilayer vesicles,'' Biophys. J.
{\bf 61} (1992) 963.}
\lref\WHH{W. Wiese, W. Harbich, and W. Helfrich, ``Budding of lipid
bilayer vesicles and flat membranes,'' J. Phys. Cond. Mat. {\bf4}
(1992) 1647.}
\lref\Sveta{S. Svetina, A. Ottova-Leitmannov\'a, and R. Glaser,
``Membrane bending energy in relation to bilayer couples concept of
red blood cell shape transformations,'' J. Theor. Biol. {\bf94} (1982)
13.}
\lref\Mich{X. Michalet, th\`ese de doctorat (Paris VII, 1994).}
\lref\Helfcz{W. Helfrich, ``Blocked lipid exchange in bilayers and its
possible influence on the shape of vesicles,'' Z. Naturforsch.
{\bf29C} (1974) 510.\optional{[invents ADE model; `the spontaneous
curvature of a bilayer around a vesicle is zero if the two halves
including the adjacent aqueous phases are in thermodynamic equilibrium']}}%
\lref\Tombook{P. Chaikin and T. Lubensky, {\sl Principles of condensed
matter physics} (Cambridge, 1994).}%
\lref\Sambook{S. Safran, {\sl Statistical thermodynamics of surfaces,
interfaces, and membranes} (Addison-Wesley, 1994).}%
\lref\Darcy{D. Thompson, {\sl On growth and form} (Cambridge, 1917).}%
\lref\Wimley{W. Wimley and T. Thompson, Biochemistry {\bf 30} (1991)
1702.\optional{[estimate lipid flipflop times]}}%
\lref\MoBZ{R. Bar-Ziv and E. Moses, \cmp{``Instability and `pearling'
states produced in tubular membranes by competition of curvature and
tension,''} Phys. Rev. Lett., {\bf73} (1994) 1392.}%
\lref\OYH{Ou-Yang Zhong-can and W. Helfrich, \cmp{``Bending energy of
vesicle membranes,''} Phys. Rev. {\bf A39} (1989) 5280.}
\lref\HeSe{W. Helfrich and R. Servuss, ``Undulations, steric
interaction and cohesion of fluid membranes,'' Nuovo Cimento {\bf3D}
(1984) 137.}
\lref\EvRa{E. Evans and W. Rawicz, ``Entropy-driven tension and
bending elasticity in condensed-fluid membranes,'' Phys. Rev. Lett.
{\bf64} (1990) 2094.}
\lref\Safi{S. Chiruvolu, {\it et al.}, \cmp{``A new phase of liposomes:
entangled tubular vesicles,''} Science, in press (1994).}
\lref\Bessis{M. Bessis, {\sl Living blood cells and their
ultrastructure,} (Springer, 1973) pp.~61, 170.}
\lref\Pouligny{B. Pouligny, private communication.}
\lref\Thpriv{B. Thomas, private communication.}
\lref\Dober{H.-G. Dobereiner, private communication.}

\lref\SeSch{J. Selinger and J. Schnur, ``Theory of chiral lipid
tubules,'' Phys. Rev. Lett. {\bf71} (1993) 4091.}
\lref\Cates{
M. Cates, ``The Liouville field theory of random surfaces,'' Europhys.
Lett. {\bf8} (1988) 719.\optional{[6/88]}}%
\lref\DeTa{P.G. de\thinspace Gennes and C.  Taupin, J. Phys. Chem.
{\bf86} (1982) 2294.}

%
%
\lref\EvYe{E. Evans and A. Yeung, \cmp{``Hidden dynamics in rapid
changes of bilayer shape,''} Chem. Phys. Lipids, in press.}%
\lref\EvSk{E. Evans and R. Skalak, {\sl Mechanics and thermodynamics
of biomembranes}  (CRC Press, 1980).}%
\lref\WSSZ{R. Waugh, J. Song, S. Svetina, and B. Zeks, \cmp{``Local and
nonlocal curvature elasticity in bilayer membranes by tether
formation,''} Biophys. J. {\bf61} (1992) 974.}%
\lref\BSZW{B. Bozic, S. Svetina, B. Zeks, and R. Waugh, ``Role of
lamellar membrane structure in tether formation form bilayer
vesicles,'' Biophys. J. {\bf61} (1992) 963.}%
\lref\BrMe{F. Brochard-Wyart, J.-M. di$\,$Meglio, and D. Qu\'er\'e,
``Theory of the dynamics of spreading of liquids on fibers,'' J. Phys.
France {\bf51} (1990) 293.}
\lref\BrLe{F. Brochard and J. Lennon, ``Frequency spectrum in the
flicker phenomenon in erythrocytes,'' J. Phys. France {\bf36} (1975)
1035.}
\lref\Seidyn{U. Seifert, ``Dynamics of a bound membrane,'' Phys. Rev.
{\bf E49} (1994) 3124.}
\lref\Brugro{R. Bruinsma, ``Growth instabilities of vesicles,'' J.
Phys. II France {\bf1} (1991) 995.}
\lref\Strutta{Lord Rayleigh, ``On the instability of jets,'' Proc. Lond.
Math. Soc. {\bf10} (1879) 4.}%
\lref\Struttb{Lord Rayleigh, \cmp{``On the instability of a cylinder of viscous
liquid under capillary force,''} Phil. Mag. {\bf34} (1892) 145.}%
\lref\Goren{S. Goren, ``The instability of an annular thread of
fluid,'' ?? 309.}%
\lref\Tomo{S. Tomotika, Proc. Roy. Soc. Lon. {\bf A150} (1932) 322.}%
\lref\promise{P. Nelson, T. Powers, and U. Seifert, to appear.}%
\lref\Liprev{R. Lipowsky, ``The conformation of membranes,'' Nature
{\bf349} (1991) 475.}
%
%
\lref\Messager{R. Messager, P. Bassereau, and G. Porte, ``Dynamics of
the undulation mode in swollen lamellar phases,'' J. Phys. France
{\bf51} (1990) 1329.}
\lref\bigchiral{P. Nelson and T. Powers, ``Renormalization of chiral
couplings in tilted bilayer membranes,'' J.  Phys. France II  {\bf3}
(1993) 1535.}

\lref\TCLWC{W. Cai and T.C. Lubensky, \cmp{``Covariant hydrodynamics of
fluid membranes,''} Phys. Rev. Lett. {\bf73} (1994) 1186.}
\lref\MiMo{D. Morse and S. Milner, ``Fluctuations and phase behavior of
surfactant vesicles,'' Preprint 1993.}
\lref\CaHe{P. Canham , J. Theor. Biol. {\bf26} (1970) 61; W.
Helfrich, Naturforsch. {\bf28C} (1973) 693.}
\lref\SMMS{See for example {\sl Statistical mechanics of membranes and
surfaces,} D. Nelson {\it et al.}, eds (World Scientific, 1989).}
\lref\DaLe{F. David and S. Leibler, ``Vanishing tension of fluctuating
membranes,'' J. Phys. II France {\bf 1} (1991) 959.} 
\lref\Kla{H. Kleinert, ``Thermal softening of curvature elasticity in
membranes,''
Phys. Lett. {\bf 114A} (1986) 263.\optional{[11/14/85]}}%
\lref\Can{P. Canham , J. Theor. Biol. {\bf26} (1970) 61}
\lref\Helfaa{W.
Helfrich, Naturforsch. {\bf28C} (1973) 693\optional{[introduces tilt
by analogy with lyotropic l.c.'s!]}.}
\lref\PNTRP{P. Nelson and T. Powers, Phys. Rev. Lett. {\bf69} (1992) 3409.}
\lref\MFM{G. Moore and P. Nelson, \cmp{``Measure for moduli,''} Nucl. Phys.
{\bf B266} (1986) 58.}
\lref\Polch{J. Polchinski, \cmp{``Evaluation of the one loop string path
integral,''} Commun. Math. Phys. {\bf 104} (1986) 37.}
\lref\Browicz{E. Browicz, Zbl. Med. Wiss. {\bf28} (1890) 625.}
\lref\DDK{F. David, ``Conformal field theories coupled to 2-D gravity in the
conformal
gauge,''
 Mod. Phys. Lett. {\bf A3} (1988) 1651; J. Distler and H. Kawai,
``Conformal field theory and 2-D quantum gravity,''  Nucl. Phys. {\bf
B321} (1989) 509.}
\lref\PoStro{J. Polchinski and A. Strominger, ``Effective string theory,''
Phys. Rev. Lett. {\bf67} (1991) 1681.}
\lref\Polbook{A. Polyakov, {\sl Gauge fields and strings}, (Harwood,
1987).}

Despite its vast complexity, the living world has always inspired
physical scientists with its habit of choosing simple geometrical
forms for its structures (see for example \Darcy). While the link
between real living systems and simple models generating similar
shapes is often tenuous at best, recent years have seen remarkable
progress in explaining the shapes of structures such as
normal and diseased red blood cells. While plasma membranes are
mixtures of thousands of lipids and proteins, extremely simple
artificial membranes consisting of a single lipid reproduce much of
their shape behavior in accordance with equally simple mathematical
models
\Liprev. To date most work has focused on equilibrium shapes, but of
course biological systems are usually not in equilibrium (\ie, dead),
and so both theoretical and experimental work has recently turned to
dynamical shape problems.

Quasicylindrical shapes are also abundant in Nature, though they have
been less studied than the bag-like shapes reminiscent of blood cells.
For example, long ago Thompson remarked~\Darcy\ a family resemblance
between certain foraminifera and the shapes of constant mean curvature
found by Delaunay~\Bouasse. These shapes are a ``peristaltic''
modulation of a cylinder, \ie\ a periodic change in its diameter. More
reasonably perhaps, Deuling and Helfrich called upon these shapes to
explain the observed `myelin figures' found inside and outside aged red
blood cells~\DeHea. Such modulated cylinders have also been seen in
several recent experiments, for example refs.~\MoBZ\Safi
; a more extreme form consisting of large pearls on a string exists as
well \MoBZ\Thpriv.

We will focus on the beautiful work of Bar-Ziv and Moses, who excited
long cylindrical lipid bilayer vesicles using laser tweezers~\MoBZ.
Their experiment seems unique in the degree of control over the
circumstances of excitation; time scales can readily be measured and
the role of thermal fluctuations is clearly seen via video microscopy.
Briefly the observed phenomenon of interest to us is as follows.
Initial preparation of the system yields stable tubular structures
with a wide variety of radii $R_0$ between 0.3--5\micron. The tubes
are
nearly straight cylinders some hundreds of microns long, anchored
at both ends by large globules of lipid. Each tube consists of a
single bilayer of DMPC or DGDG. The high temperature used
($\sim45^\circ$C) precludes any in-plane ordering of lipid molecules,
so that away from the localized excitation a pure fluid membrane model
suffices to describe their shapes. Initially the system is somewhat
flaccid, as seen from visible thermal undulations and the fact
that the tubes are not quite straight.

Application of a laser spot localized to $\sim0.3$\micron\ produces a
dramatic {\df transformation to the peristaltic shape.}
Greater laser power is required for larger tubules.
Once formed, the peristaltic shape has
a well-defined wavelength $\la_0$ which is uniform over dozens of
wavelengths. Whatever the initial radius $R_0$, $\laz$ is found to be
$2\pi R_0/k_0$ where the dimensionless wavenumber $k_0$ is always between
0.64 and 1, and typically about $0.8$.
After prolonged tweezing
some buildup of lipid becomes visible at the point of application of
the laser. As the modulation grows
more pronounced, $k$ grows from $k_0$ to become slightly
greater than 1.  The modulated state is {\it tense}: visible thermal
fluctuations are suppressed and the tube draws itself straighter than
initially.

In this Letter we give a dynamical theory for the initial pearling
instability\optional{\ref{Some details appeared in P. Nelson, short talk
at the Workshop on Biomolecular Materials, ITP August 1994.}}.  In
particular we will show how to
compute the preferred wavenumber $k_0
$ in terms of $R_0$ and independently measured
material parameters. \clip{We will
propose a mechanism for the subsequent decrease of $\la$
and the quenching and recovery of large thermal fluctuations. }Our
model is an elaboration of the suggestion in \MoBZ\ that the
instability is of Rayleigh type, but with some significant changes to
Rayleigh's classic analysis \Struttb.
For example, the Rayleigh mechanism predicts $k_0=0$.
Our mechanism boils down to a competition between a driving force,
membrane tension induced by the laser, and
ordinary hydrodynamic drag. After a simple back-of-the-envelope
calculation we will add to our picture the bending and stretching
moduli for the bilayer membrane as well as the friction between its
two leaves. On short time scales
the friction tends to lock the layers together~\Evansb\SeLa, leading
to a transient
effective spontaneous curvature. Using independently measured values
for all the material parameters we will then get $k_0$ in rough agreement
with experiment. More details will appear elsewhere~\promise.

We cannot simply identify the modulated
cylinders as equilibrium surfaces of constant mean curvature, since
nothing in the problem selects a curvature. Unlike ref.~\DeHea\ we
have no {\it chemical} asymmetry between the inner and outer fluids to
generate a tendancy to bend, and hence no spontaneous curvature.
Indeed the tubules are polydisperse, each
one's radius being set by the distance $L_0$ between terminal globules
and the volume which happened to get trapped during formation.
Nor do we have a fixed {\it pressure} difference as imagined in
\Darcy, which could have set a curvature by a Laplace-type law.
Nor can we appeal to an
area-difference {\it strain} frozen in at formation (unlike closed
vesicles \EvHe\MiaSeif\MSWD); each layer of the
tubule is initially in equilibrium with a common reservoir, the
terminal globule, and this implies that the effective spontaneous
curvature vanishes~\EvHe.\foot{Actually this is advantageous, as our
results do not depend on an unobservable parameter varying from one
tubule to the next.}
Finally, the Delaunay shapes \Bouasse\Darcy\ have an initial
instability at wavelength $k=1$ \promise, larger than any observed
value. In fact we cannot understand the observed shapes as equilibrium
shapes for some suddenly modified elastic energy, so we turn to dynamics.

The most famous dynamic instability in cylindrical geometry is that of a
column or jet of water in air \Strutta. Despite a superficial
resemblance to the pearling instability, however, Rayleigh's original
analysis does not apply to the micron scale, where water is viscid.
Following Plateau \ref\Plateau{J. Plateau, {\sl Statique experimentale
et theorique des liquides \cmp{soumis aux seules forces moleculaires}}
(Gautier-Villars, 1873).}, Rayleigh later
showed that for a cylinder of viscid fluid (a `thread of treacle') in
air, interfacial tension gives a fastest-growing
unstable mode at $k\max=0$ \Struttb, which is far from the observed
$k_0$. In fact Rayleigh's answer depends
on a boundary condition which was appropriate for his problem but not
for ours; imposing no-slip boundary conditions
at a nearly incompressible membrane instead of Rayleigh's condition of no
tangential force will give us $k\max\not=0$ (see also \Brugro). We
will then justify this simple
model by incorporating the full elastic and dynamic structure of the
membrane.

To get started let us consider the nature of the excitation.
Initially our membrane is under almost zero tension, as seen from the
thermal motion. When the laser comes close to the membrane, nothing
happens: local heating is not important. When the laser spot {\it
touches} the membrane, it pulls material in by the usual tweezer
effect. While it is hard to calculate the exact tension so
induced, we may easily estimate it as follows\foot{We thank R.
Bruinsma for suggesting this estimate. Any heating caused by direct
contact with the membrane will simply increase slightly the area in
the trap without changing the desire of molecules outside the trap to
get in.}: the applied laser power of
$\sim50$mW, spread over a spot of diameter $0.3$\micron, corresponds
to an energy density in vacuum $\CE$ of $3\cdot10^4$\erg cm$^{-3}$.
Taking
the dielectric contrast between water and lipid at this frequency to
be of order $\delta\epsilon=0.23$
\ref\Isrbook{J. Israelachvili, {\sl Intermolecular and surface forces}
(Academic Press, 1992).}%
, we see that when a lipid
molecule falls into the trap displacing water we gain an energy
\ref\JDJ{J. D. Jackson, {\sl Classical Electrodynamics} (Wiley, 1975)
pp. 107, 160.}{}
$\sim\CE\dl
\ep\cdot a_0D$, where $a_0$ is the area of the molecule's head and $D$ is
its total length. Taking $2D\sim40$\AA, we get that each unit of
bilayer area sucked into the trap yields an energy gain of
$\Sigma\sim10^{-3}$\erg\cmmt. While this figure is probably an
overestimate, we see that the trap generates a tension well in excess
of the critical value~\MoBZ\ $\Sigma\crit\sim{\kappa\over
R_0^2}\sim10^{-4}$\erg \cmmt\ needed to trigger shape transformations,
where $\kappa\sim0.6\eem{12}$\erg\ is the bending stiffness of DMPC
bilayers and we took $R_0=0.7$\micron\ for illustration.
Let $\sigma\equiv\Sigma R_0^2/\kappa$ denote the dimensionless
tension; thus $\sigma\sim10$. Following \MoBZ\ we will take this
tension to be distributed uniformly over the whole surface, even
though in fact it propagates outwards from the laser spot.
If the laser is removed the tension reverts
back to zero,
thermal fluctuations resume, and the tubule relaxes back to its
initially stable cylindrical shape, as observed.

We work in cylindrical coordinates and describe our shape by the locus
$\rhoPN= R_0(1+u(z,t))$ where $|u|\ll 1$. (Non-axisymmetric
perturbations turn
out to be stable \promise.) For our linearized analysis we
can treat each Fourier mode separately, except the constant term
$u_0$. Since $u_0$ is the only mode which can decrease the volume,
volume conservation in an infinite cylinder requires  (see
\promise) that we
choose $u(z)=-(\ux)^2+2\ux\cos(kz/R_0)$, where we truncate all
formulas to $\CO(u^2)$ and $\ux$ is a function of time which we are to
find. Using the area element \OYH\ $\dd S=[1+ u+\half R_0^2(\nabla
u)^2]\, R_0\dd z\dd\phi$, we at once see that this perturbation decreases
the area $A=\int\dd S=A_0[1+(k^2-1)(\ux)^2]$ only for
$k<1$~\Plateau. Hence membrane tension cannot destabilize modes with
$k>1$, as observed. For $k<1$, the laser does work $-\dl
A\cdot\Sigma$ on our system as the modulation $u_k$ grows.

Where does this energy go? On micron scales we may ignore
kinetic energy, but some energy will go to viscous dissipation inside
and outside the tube. As mentioned, energy can also go into the
internal structure of the bilayer, for example the bending elasticity,
but let us neglect such
complications for our first estimate. As our vesicle changes
shape, conservation requires a central flow velocity $v_z(\rhoPN=0)\sim
(R_0/k)\dot u_k$ to transport the fluid from the troughs
to the crests. But membrane
incompressibility and no-slip boundary conditions between the layers
and adjacent fluid require a much smaller value of $v_z(r=R_0)\sim0$ at the
boundary. Thus we get a velocity gradient and a shear dissipation
of $2A_0R_0\eta\LambdaPN(k)\inv\dot u_k^2$, where $\eta$ is the
viscosity of water and the dynamical factor
$\LambdaPN (k)$ is proportional to $k^2$ at small $k$. Equating this
power loss to  the gain $-\dot A\Sigma$ yields a growth rate
\eqn\enagro{\gamma_k\equiv\dot u_k/u_k=\Sigma\LambdaPN(k)(1-k^2)
/R_0\eta\quad.}
Approximating $\LambdaPN(k)$ by its small-$k$ (Poiseuille) form, we
see that $\gamma$ reaches its maximum at $k\max=1/\sqrt2$, right in
the experimentally observed range. A more exact solution of
viscid hydrodynamics inside and outside a moving boundary
with incompressible-layer boundary conditions gives \promise
\eqn\eUdo{\LambdaPN(k)=-{1\over 2k}{
[k(K_0^2-K_1^2)+2K_0K_1][k(I_0^2-I_1^2)-2I_0I_1]   \over
2I_0K_0/k + k(I_1^2K_0^2-I_0^2K_1^2)
}
\quad,}
where $I_\nu, K_\nu$ are the usual Bessel functions \JDJ, evaluated at
$k$. With this $\LambdaPN$ we get $k\max=0.68$. Note that $k\max$ is a
purely geometric constant because we cannot form any length scale from
the tension and viscosity. We also see from \enagro\ that the tension
$\Sigma$, and hence laser power, needed to get noticeable growth rate
increases with tubule radius $R_0$, as observed \MoBZ.

We now outline how to account for the internal dynamics of the
membrane, and when these will be important to our problem. Symmetric
bilayers resist bending and stretching with an elastic energy which we
may write as \SeLa\promise
\eqn\efone{F[H,\phi^+,\phi^-]=\int\dd S{\kappa\over2}\left\{
(2H)^2+{\beta\over2d^2}\Bigl[\bigl({\phi^+\over\phi_0}-1\bigr)^2
+\bigl({\phi^-\over\phi_0}-1\bigr)^2\Bigr]\right\}\quad.
}
Here $H$ is the mean curvature of the bilayer midplane and $\phi^\pm$
are the lipid densities
of outer (respectively inner) monolayers, measured at the neutral
surfaces of each monolayer.
$\kappa$ is the usual static bilayer bending stiffness, {\df
$d\sim D/2$ is the distance from the
bilayer midplane to the monolayer neutral surface~\WSSZ%
,} and
$\beta\equiv Kd^2/\kappa$, where $K$ is the bilayer compression
modulus. For DMPC, $\beta\sim3.5$ \MSWD. We have dropped the
topological term as usual. In equilibrium $\phi^\pm$ can adjust to
their preferred values and we recover the usual curvature model. More
generally, let us rephrase \efone\ in terms of the densities
$\psi^\pm=\phi^\pm(1\mp 2Hd)$ referred to the bilayer midplane.
Initially we have $\psi^\pm=\psi_0^\pm\equiv\phi_0(1\mp2H_0d)$, but after
excitation $\psi^\pm$ can change along with $H$. Letting $\rho^\pm$ be
the relative density change, $\rho^\pm=\psi^\pm/\psi_0^\pm-1$, the
density term of \efone\ becomes
$\lfr K4\bigl[(\rho^++2(\dl H)d)^2+(\rho^--2(\dl H)d)^2\bigr]$ where
$\dl H=H-H_0$ is the change of curvature.

Besides hydrodynamic drag and elasticity, there is one more sink of
energy, the interlayer
friction. To estimate the
importance of this resistance, write  the frictional force per unit
area as $b(\dvt)$ where $b$ is a constant and
$\narrowtilde v_\pm$ are the tangential layer velocities
\Evansb\EvYe\SeLa.  {\df Then} \eqn\erone{R_1\equiv
bd^2/\eta} is a new crossover
length scale in the problem \SeLa. For length scales much smaller than
$R_1$, bilayer friction can dominate hydrodynamic dissipation unless the
former somehow vanishes.
In this high-friction regime the hydrocarbon chains temporarily lock
together, so that under a sudden disturbance the membrane acts like a
thin plate, remembering its initial curvature.%
\foot{\optional{The required extra area/length for a Delaunay surface
may come from the observed straightening of the tubes in the nonlinear
regime. }The reader may wonder how to reconcile this effect with the
appearance of very thin tethers in the extreme nonlinear regime \MoBZ.
Note that while these tethers have greatly increased curvature, still
their {\it area} is so small that frictional losses incurred in their
formation are not too great. We also note in passing that the radius
of the long tethers in ref.~\Evansb\ is related to the tension by
$\Sigma=\kappa/R_{\rm tether}^2$.
Substituting the tension used here into this relationship gives a
tether radius roughly comparable to that seen in \MoBZ.}
Both diffusion measurements \ref\MSE{R.
Merkel, E. Sackmann, and E. Evans, J. Phys. (Paris) {\bf50} (1989)
1535.}\ and tether-pulling experiments \Evansb\  yield $bd^2\sim
10^{-6}$\erg sec\thinspace cm$^{-2}$ while the viscosity
$\eta\sim 10^{-2}$\erg sec\thinspace cm$^{-3}$, 
so we get $R_1\sim 1$\micron, comparable to our system's size $R_0$.

The projected density fluctuations $\rho^\pm$ are related to
$\narrowtilde v_\pm$ by conservation\foot{In general, conservation
laws on moving curved surfaces are tricky (see
\ifx\prlmode\testp{}
W. Cai and T.C. Lubensky, Phys. Rev. Lett. {\bf73} (1994) 1186),
\else\TCLWC),
\fi
but for our
linearized analysis this treatment suffices.}, $\nabla_\parallel
\cdot\narrowtilde v_\pm=-\dot\rho^\pm$. While friction thus obstructs
changes in $\rho^+-\rho^-$, nothing prevents  $\rho^++\rho^-$ from
quickly adopting its preferred value of zero and we finally obtain the
quadratic approximation to the energy
\eqn\eftwo{F[H,\hat\rho]={\kappa\over2R_0^2}\int\dd S
\left[\CP(k^2)u^2+\lfr\beta4 \hat\rho^2+\beta(1-k^2)\hat\rho u\right]
\quad,}
where $\hat\rho\equiv(R_0/d)(\rho^+-\rho^-)$ and
\eqn\eP{\CP(k^2)\equiv\lfr32
-\sigma+(\sigma-\half)k^2+k^4+\beta(1-k^2)^2\quad.}
Normal force balance now sets $T_{\rhoPN\rhoPN}^+-T_{\rhoPN\rhoPN}^-
=R_0\inv\cdot\dl F/\dl u$, where $T_{ij}^\pm$ are the fluid stress
tensors outside (respectively inside) the vesicle, evaluated at the
boundary. In linear approximation we have
$T_{\rhoPN\rhoPN}^+-T_{\rhoPN\rhoPN}^-=-\eta\LambdaPN(k)\inv\dot u$,
where $\LambdaPN(k)$ is the factor quoted in eqn.~\eUdo.%
\foot{We may
neglect the small boundary velocities $\narrowtilde v^\pm\sim\dot
ud\ll\dot uR_0$ in
this part of the calculation \promise.}

To solve for both $u(z,t)$ and $\hat\rho(z,t)$ we must supplement the
normal force-balance equations by the difference between the two
tangent force-balance equations \SeLa: $-b(\narrowtilde v^\pm -
\narrowtilde v^\mp)=\nabla_\parallel(\dl F/\dl\rho^\pm)$. We have
retained only interlayer friction on the left side, since it turns out
to dominate both the 2d viscosity \EvSk\SeLa\promise\
and the traction $\pm T^\pm_{rz}$ of the fluid \SeLa\promise.
Subtracting the two equations and using the lipid conservation
equations 
gives us $\pa\hat{\rho}/\pa t=-(k^2/bd^2)\,\dl F/\dl\hat\rho$, where again
$\hat \rho\equiv(R_0/d)(\rho^+-\rho^-)$. We combine the force balance
equations using \eftwo\ to get
\eqn\eyuck{\pd{}t\pmatrix{ u\cr\hat{\rho}\cr}=
-{\kappa\over R_0^3\eta}\pmatrix{\LambdaPN(k)&\cr&k^2R_0/R_1\cr}
\pmatrix{\CP(k^2)& \half\beta(1-k^2)\cr
         \half\beta(1-k^2) & \beta/4 \cr}
\pmatrix {u\cr\hat{\rho}\cr}
\quad.}
The desired growth rate is the positive eigenvalue, if any, of this
linear problem, \ie\ $\ga\approx(-180\,{\rm sec}\inv)\cdot\la$, where
we took a typical $R_0=0.7$\micron\  and $\lambda$ solves $\lambda^2-
\lambda(g_1+y)+g_0y=0$. Here $g_0=\LambdaPN(k)[\lfr32 -\sigma+
k^2(\sigma-\half)+k^4]$ is the growth rate with zero friction and
$g_1=g_0+\LambdaPN(k)\beta(1-k^2)^2$ is the growth rate at infinite
friction; $y\equiv\beta R_0k^2/4R_1$.

Using measured values of the parameters quoted above, \eyuck\ with
\eUdo\ yields a broad distribution of growth rates peaked around
$k\max\approx 0.65$ for any tension $\sigma$ near our estimate
$\sigma\sim 10$, which is consistent with some of the observations and
certainly better than the $k\max\sim0$ found in Rayleigh's problem
\Struttb. (Alternately the value of $\sigma$ can be inferred from the
observed growth rate $\ga\max$, once this is measured.)
We also see why the naive treatment we gave first is
so accurate: for large tension we have
$g_0\approx g_1$ and the growing eigenvalue
$\la\approx\sigma(k^2-1)\LambdaPN(k)$ becomes completely insensitive
to the friction. {\df Physically, at large tension both the bending
stiffness and the transient spontaneous curvature
due to the layer friction become unimportant.} To obtain the shorter
wavelengths seen in other observations we must either suppose that in
those cases the tubule had already passed into its nonlinear regime or
that the effective values of the friction coefficient $b$ and the
area-difference elasticity parameter $\beta$ are somehow larger than
our estimates. Of course there may be still more physics which we have
missed. For example we have not attempted here to study the
propagation of the disturbance from a point source.

In short, direct application of laser tweezers can serve as a
tool for exploring the dynamics of membranes by triggering
visible shape transformations, complementary to other controlled
techniques such as tether-pulling.
We have seen how the laser-membrane interaction may be
modeled by a very simple mechanism. We found that the membrane acts as
a nearly incompressible boundary for the surrounding solvent, which
significantly changes the instability of cylindrical geometry
{}from the Rayleigh case in ways which have been observed. Analysis
similar to that presented here should also prove useful in
understanding other dynamical problems involving shape changes of
vesicles, and in particular problems at lower tension, where the
internal membrane dynamics will play a larger role.
One may hope that the insight
thus gleaned can find uses in understanding
processes, such as vesiculation, of more direct biological interest.


{\frenchspacing
\ifx\prlmode\testp\else\vskip1truein \leftline{\bf Acknowledgements}
\noindent\fi
We would like to thank
R. Bruinsma,
F. David,
B. Fourcade,
M. Goulian,
T. Lubensky,
D. Nelson,
X. Michalet,
S. Safran,
and especially R. Bar-Ziv and E. Moses for countless discussions.
P.N.\ also thanks CEA Saclay, the Weizmann Institute, and the Institute
for Advanced Study, and P.N.\ and U.S. thank ITP Santa Barbara for
their hospitality and partial support, while this work was done.
PN acknowledges NSF
grants PHY88-57200
and PHY89-04035, 
and the Donors of the Petroleum Research Fund,
administered by the American Chemical Society, for the partial support
of this research.}

\footatend\bigskip\immediate\closeout\rfile\writestoppt
  \centerline{{\bf References}}\bigskip{\frenchspacing%
  \parindent=20pt\escapechar=` \input refs.tmp\vfill\eject}\nonfrenchspacing

\bye